\documentclass[epj]{webofc}
\usepackage[utf8]{inputenc}
\usepackage[varg]{txfonts}   % Web of Conferences font
\usepackage{booktabs}
\usepackage{xcolor}
\usepackage{multicol}

\usepackage{algorithm2e}
\SetAlFnt{\scriptsize}
\SetAlCapFnt{\tiny}
\SetAlCapNameFnt{\tiny}

\definecolor{darkred}{rgb}{0.4,0.0,0.0}
\definecolor{darkgreen}{rgb}{0.0,0.4,0.0}
\definecolor{darkblue}{rgb}{0.0,0.0,0.4}
\usepackage[bookmarks,linktocpage,colorlinks,
    linkcolor = darkred,
    urlcolor  = darkblue,
    citecolor = darkgreen]{hyperref}
\usepackage{subfigure}
\wocname{EPJ Web of Conferences}
\woctitle{Lattice2017}
\begin{document}
%%%%%%%%%%%%%%%%%%%%%%%%%%%%%%%%%%%%%%%%%%%%%%%%%%%%%%%%%%%%%%%%%%%%%%%%%%%%%
%
\selectlanguage{english}
%----------------------------------------------------------------------------
\title{%
Baryonic and mesonic 3-point functions with open spin indices
}
%----------------------------------------------------------------------------
\author{%
\firstname{Gunnar S.} \lastname{Bali}\inst{1,2} \and
\firstname{Sara} \lastname{Collins}\inst{1} \and
\firstname{Benjamin} \lastname{Gl\"aßle}\inst{1} \and
\firstname{Simon} \lastname{Heybrock}\inst{3} \and
\firstname{Piotr} \lastname{Korcyl}\inst{1,4} \and
\firstname{Marius}  \lastname{L\"offler}\inst{1}\fnsep\thanks{Speaker,
\email{marius.loeffler@physik.uni-regensburg.de}} \and
\firstname{Rudolf} \lastname{R\"odl}\inst{1} \and
\firstname{Andreas} \lastname{Sch\"afer}\inst{1}
% etc.
}

%----------------------------------------------------------------------------
\institute{%
Institut f\"ur Theoretische Physik, Universit\"at Regensburg, D-93040 Regensburg, Germany
\and
Department of Theoretical Physics, Tata Institute of Fundamental Research, Homi Bhabha Road, Mumbai 400005, India
\and
European Spallation Source, 225 92 Lund, Sweden
\and
M. Smoluchowski Institute of Physics, Jagiellonian University, ul. \L ojasiewicza 11, 30-348 Krak\'ow, Poland
}
%----------------------------------------------------------------------------
\abstract{%
We have implemented a new way of computing three-point correlation functions. It is based on a factorization of the entire correlation function into two parts which are evaluated with open spin- (and to some extent flavor-) indices. This allows us to estimate the two contributions simultaneously for many different initial and final states and momenta, with little computational overhead. We explain this factorization as well as its efficient implementation in a new library which has been written to provide the necessary functionality on modern parallel architectures and on CPUs, including Intel’s Xeon Phi series.
}
%----------------------------------------------------------------------------
\maketitle
%----------------------------------------------------------------------------
\section{Introduction}\label{intro}

Observables constructed with the use of three-point correlation functions can describe a multitude of physical phenomena, such as the parton structure of hadrons or their weak transitions, depending on the operator type at the insertion. The relevant strong interaction matrix elements can be computed using Lattice Quantum Chromodynamics. Traditionally, a method employed to that aim was the sequential source method \cite{martinelli:1989}. The numerical cost of this is high because a new inversion is necessary for each final momentum at the sink. In this contribution we present a stochastic algorithm which circumvents this limitation and disentangles the number of inversions of the lattice Dirac operator from the number of sink/insertion momenta. The implementation we propose parallelizes the computations in such a way that multiple source positions and multiple insertion positions can be estimated simultaneously. Moreover, by storing the uncontracted data, with all spin indices open, on disk, we enable the user to analyse any channel of interest at a later stage.

An aspect of our implementation which we wish to highlight in this contribution is the extensive use of vectorization. As the compute power of today's processors relies on longer and longer vector registers of additional vector processing units, an adequate data layout is required to efficiently use these resources. We explain our data layout which supports large vector registers, therefore enabling us to efficiently run our code on the Intel's Xeon Phi processors featuring 512 bit AVX vector instructions. 

Our implementation is based on a framework developed specifically for Intel's Xeon Phi processors, called \texttt{LibHadronAnalysis} see, e.g., \cite{lha:2017, heybrock:2015}. This library contains additional routines for computing meson and baryon spectra, meson and baryon distribution amplitudes and many other objects. Compared to a naive implementation in Chroma \cite{Edwards:2004} using QDP++ \cite{Edwards:2004} objects it provides speed-up factors of the order 10-20 on the KNC and KNL architectures.

Below we describe in detail the algorithm. In particular we explain how the computation of a three-point correlation function can be separated into two, largely independent parts, the ``spectator'' and the ``insertion'' parts. We pay special attention to the parallelization schemes which are different for each of these parts. Subsequently, we present benchmarks showing the performance of our implementation on a KNL cluster, and then we conclude.

%----------------------------------------------------------------------------
\section{Stochastic baryonic and mesonic three-point correlation functions}\label{sec-1}

In this section we introduce the three-point correlation function we are interested in and show how it can be factorized into two, largely independent parts. We only show explicit formulae for the case of meson three-point functions, for the sake of notational simplicity. A similar approach was used in Refs. \cite{evans:2010, Alexandrou:2013xon, Bali:2013gxx, Yang:2015zja}, however, here we keep all spin indices open. 

\subsection{General structure}

\begin{figure}[thb]
  \centering
  \includegraphics[width=8cm,clip]{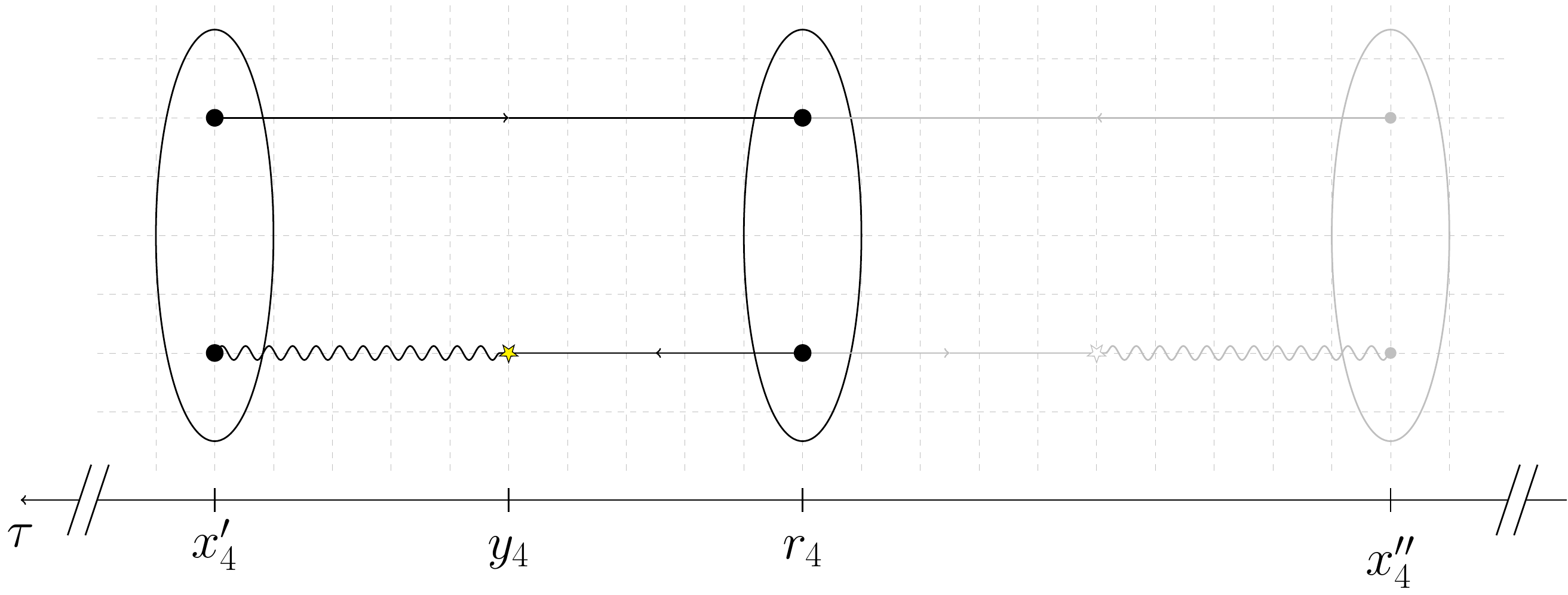}
  \caption{Sketch of the structure of a generic three-point correlation function.}
  \label{fig-6}
\end{figure}

A three-point meson correlation function (c.f., figure \ref{fig-6}) with a source $C(r)$, a sink $A(x')$ and an insertion operator $I(y)$, located at timeslices $r_4$, $x'_4$ and $y_4$ respectively, reads

\begin{align}
\langle A(x^\prime) \, I(y) \, C(r) \rangle &=  \textrm{tr} \left[ G_{f_1}(r,x^\prime) \,\, \Gamma_{\text{snk}} \,\, G_{f_2}(x^\prime,y) \,\, \Gamma_{\text{ins}} \,\, G_{f_3}(y,r) \,\, \Gamma_{\text{src}} \right] \nonumber \\
&= \,\, \delta_{ab} \,\, \delta_{a'b'} \,\, \delta_{\tilde{a} \tilde{b}} \,\,  \Gamma_{\text{snk}}^{\alpha' \beta'} \,\, \Gamma_{\text{ins}}^{\tilde{\alpha} \tilde{\beta}} \,\, \Gamma_{\text{src}}^{\beta \alpha} \,\, G_{f_1}(r,x^\prime)^{\alpha \alpha'}_{a a'} \,\, G_{f_2}(x^\prime,y)^{\beta' \tilde{\alpha}}_{b' \tilde{a}} \,\, G_{f_3}(y,r)^{\tilde{\beta} \beta}_{\tilde{b} b} 
\end{align}
where
\begin{align}
%\textnormal{Annihilator: } 
A(\mathbf{x^{\prime}},x^{\prime}_4) &= \delta_{a' b'} \,\, \overline{\psi}_{f_2}(\mathbf{x^{\prime}},x^{\prime}_4)_{a'}^{\alpha'} \,\, \Gamma^{\alpha' \beta'}_{\text{snk}} \,\, \psi_{f_1}(\mathbf{x^{\prime}},x^{\prime}_4)_{b'}^{\beta'} \,\, , \label{an} \\
                % F * F_bar has to be a Lorentz scalar, therefore use the definition that F_bar := F_star * gamma_4
%\textnormal{Creator: } & 
C(\mathbf{r},r_4) &= \delta_{ba} \,\, \overline{\psi}_{f_1}(\mathbf{r},r_4)_b^{\beta} \left( \gamma_4 \Gamma^{\dagger} \gamma_4 \right)^{\beta \alpha} \,\, \psi_{f_3}(\mathbf{r},r_4)_a^{\alpha} \,\, , \label{cr} \\
%\textnormal{Current: } 
I(\mathbf{y},y_4) &= \delta_{\tilde{a} \tilde{b}} \,\, \overline{\psi}_{f_3}(\mathbf{y},y_4)_{\tilde{a}}^{\tilde{\alpha}} \,\, \Gamma_{\text{ins}}^{\tilde{\alpha} \tilde{\beta}} \,\, \psi_{f_2} (\mathbf{y},y_4)_{\tilde{b}}^{\tilde{\beta}} \,\, , \label{in} 
\end{align}
are the annihilation, creation and insertion operators respectively, with  $f_i \in \{l,s,c\}$ (light, strange, charm). $G_{f_i}(r,x)$ is a standard fermion propagator from $x$ to $r$ of flavor $f_i$. We use the convention to denote the annihilation spin and color operator indices with primed Greek and Latin letters $\alpha'$, $a'$, creation operator indices with ordinary Greek and Latin letters $\alpha$, $a$, and the insertion operator indices with tilde Greek and Latin letters $\tilde{\alpha}, \tilde{a}$. $\Gamma_{\text{ins}}$ can contain local derivatives and $A$ and $C$ may contain quark smearing.

At this point we replace one of the propagators by its stochastic estimate
\begin{equation}
G_{f_2}(y, x')^{\tilde{\alpha} \beta'}_{\tilde{a} b'} \approx \frac{1}{N} \sum_{i=1}^{N} \,\, s_{i, \, f_2}(y)^{\tilde{\alpha}}_{\tilde{a}} \,\, \left(\eta_i^{*} \right)(x')^{\beta'}_{b'} \,\, ,  
\end{equation} 
where the sum runs over $N$ realizations of the noise source vector $\eta_i(x')$, with the properties
\begin{align}
\frac{1}{N} \, \sum_{i=1}^{N} \,\, \left(\eta_i \right)(x)^{\alpha}_a &= 0 + \mathcal{O} \left(\frac{1}{\sqrt{N}} \right)\,\, ,\\
\frac{1}{N} \, \sum_{i=1}^{N} \,\, \left(\eta_i \right)(x)^{\alpha}_a \left(\eta_i^{*} \right)(x^{\prime})^{\alpha'}_{a'} &= \delta_{xx'} \,\, \delta_{\alpha \alpha'} \,\, \delta_{aa'} + \mathcal{O} \left(\frac{1}{\sqrt{N}} \right) \,\, .
\end{align}
The $\left(\eta_i \right)(x)$ are time partitioned and set to zero, unless $x_4 = x'_4$ or $x_4 = x''_4$.

In figure \ref{fig-6} we show the generic structure of a three-point correlation function in the meson case. The central ellipse denotes a meson created with operator Eq.~(\ref{cr}) in the middle of the lattice at timeslice~$r_4$. The meson is annihilated by the operator from Eq.~(\ref{an}) at the timeslice $x'_4$ as depicted by the left-most ellipse. Arrows denote exact point-to-all propagators. The star at timeslice $y_4$ denotes one of the possible positions of the insertion operator. The wiggly line is used to plot the stochastic all-to-all propagator. We call these four elements the ``forward'' correlation function as opposed to the shaded mirror-reflected graph on the right hand side which corresponds to the ``backward'' process. The forward and backward diagrams are estimated simultaneously which allows for increased statistics. 

The introduction of the stochastic propagator allows us to factorize the correlation function $\langle A(x^\prime) \, I(y) \, C(r) \rangle$ into two parts \cite{evans:2010} as follows
\begin{multline}
\langle A(x^\prime) \, I(y) \, C(r) \rangle = \,\, \delta_{ab} \,\, \delta_{a'b'} \,\, \delta_{\tilde{a} \tilde{b}} \,\,  \Gamma_{\text{snk}}^{\alpha' \beta'} \,\, \Gamma_{\text{ins}}^{\tilde{\alpha} \tilde{\beta}} \,\, \Gamma_{\text{src}}^{\beta \alpha} \,\, \times \\ 
\times  \underbrace{ \left[ \gamma_5 \,\, G_{f_1}^{\dagger}(x^\prime,r) \,\, \gamma_5 \right]^{\alpha' \alpha}_{a' a} \,\, \left[ \eta_i(x^\prime) \,\, \gamma_5 \right]^{\beta'}_{b'} }_{\hat{=}\,\textnormal{Spectator}}\,\, \underbrace{ \left[ \gamma_5 \,\, s_{i, \, f_2}(y) \right]^{*\,\,\tilde{\alpha}}_{\,\,\,\,\, \tilde{a}} \,\, G_{f_3}(y,r)^{\tilde{\beta} \beta}_{\tilde{b} b}}_{\hat{=}\,\textnormal{Insertion}} 
\end{multline}
We define the spectator $S_{i, \, f_1}(\mathbf{p},x^\prime_4)^{\, \beta' \, \alpha' \,\alpha}_{a}$ and the insertion 
$I_{i, \, f_2, \, f_3 }(\mathbf{q},y_4)^{\tilde{\alpha} \, \tilde{\beta} \, \beta}_{a} $ parts 
\begin{align}
S_{i, \, f_1}(\mathbf{p'},x^\prime_4)^{\beta' \, \alpha' \, \alpha}_{a} &= 
\sum_{\mathbf{x^\prime}} \,\, \delta_{a' b'} \,\, \left[ \eta_i(x^\prime) \,\, \gamma_5 \right]^{\beta'}_{b'} \left[ \gamma_5 \,\, G_{f_1}^{\dagger}(x^\prime,r) \,\, \gamma_5 \right]^{\alpha' \alpha}_{a' a} \cdot e^{-i \mathbf{p'} \cdot \mathbf{x^\prime}} \,\, ,\label{spectator}\\
I_{i, \, f_2, \, f_3}(\mathbf{q},y_4)^{\tilde{\alpha} \, \tilde{\beta} \, \beta}_{a} &= \sum_{\mathbf{y}} \,\, \delta_{ab} \,\, \delta_{\tilde{a} \tilde{b}} \,\, \left[ \gamma_5 \,\, s_{i, \, f_2}(y) \right]^{*\,\,\tilde{\alpha}}_{\,\,\,\,\, \tilde{a}} \,\, G_{f_3}(y,r)^{\tilde{\beta} \beta}_{\tilde{b} b} \cdot e^{i \mathbf{q} \cdot \mathbf{y}} \,\, , \label{insertion}
\end{align} 
where we have assumed $\mathbf{r} = 0$. Otherwise we have to replace $\mathbf{x}' \rightarrow \mathbf{x}' - \mathbf{r},\,\, \mathbf{y} \rightarrow \mathbf{y} - \mathbf{r}$.

\subsection{Spectator part}

\begin{figure}[thb]
  \centering
  \includegraphics[width=8cm,clip]{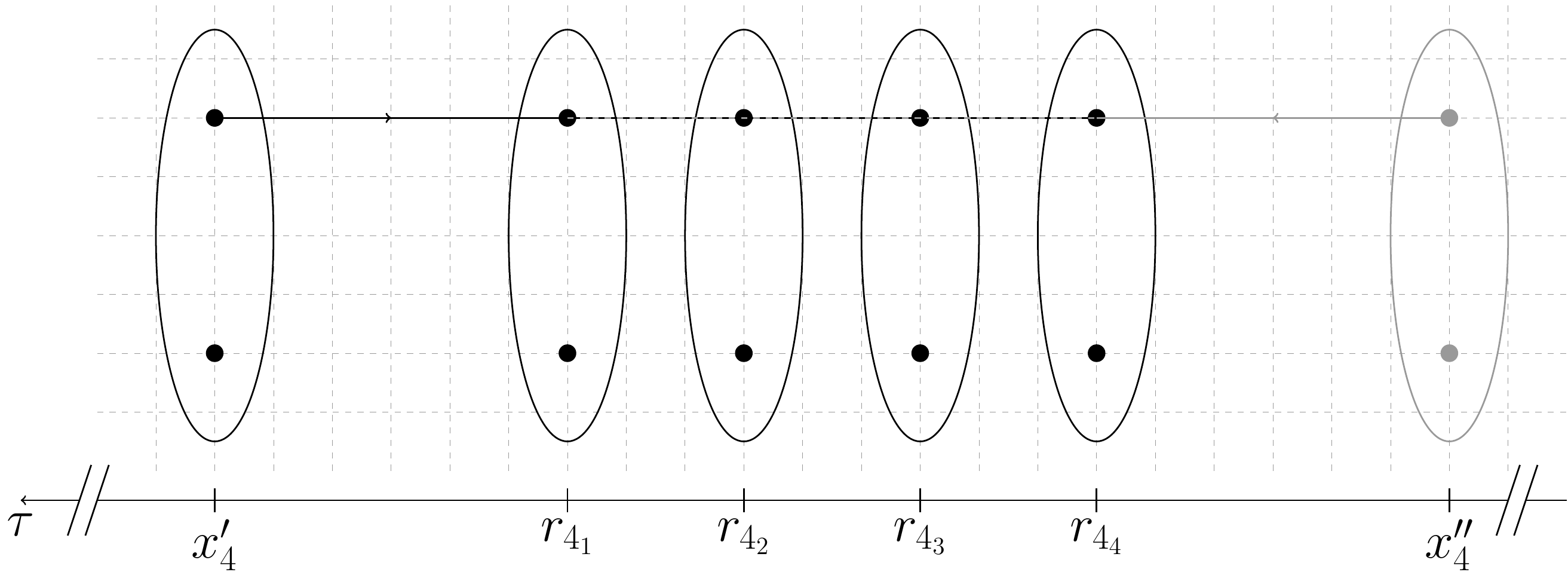}
  \caption{Parallelization of the spectator part of the three-point correlation function. Propagators sourced at different timeslices denoted by different
solid ellipses are redistributed among all MPI ranks so that each rank has at least one set of propagators to work with.}
  \label{fig-5}
\end{figure}

The computation of the spectator part consists of the contractions of propagators at the timeslices where the source and the sink are located. Naively only the MPI ranks working on the timeslices $r_4$ and $x_4'$, $x''_4$ would work. In our implementation we prepare a set of propagators sourced from different temporal source positions, as shown on figure \ref{fig-5}. We redistribute these propagators among the different MPI ranks in such a way that each rank has at least one propagator. The computation of the spectator part for each source position is than performed simultaneously. The Fourier transformation Eq.~(\ref{spectator}) fixes the momentum $\mathbf{p'}$ at the sink.

\subsection{Insertion part}

\begin{figure}[thb]
  \centering
  \includegraphics[width=8cm,clip]{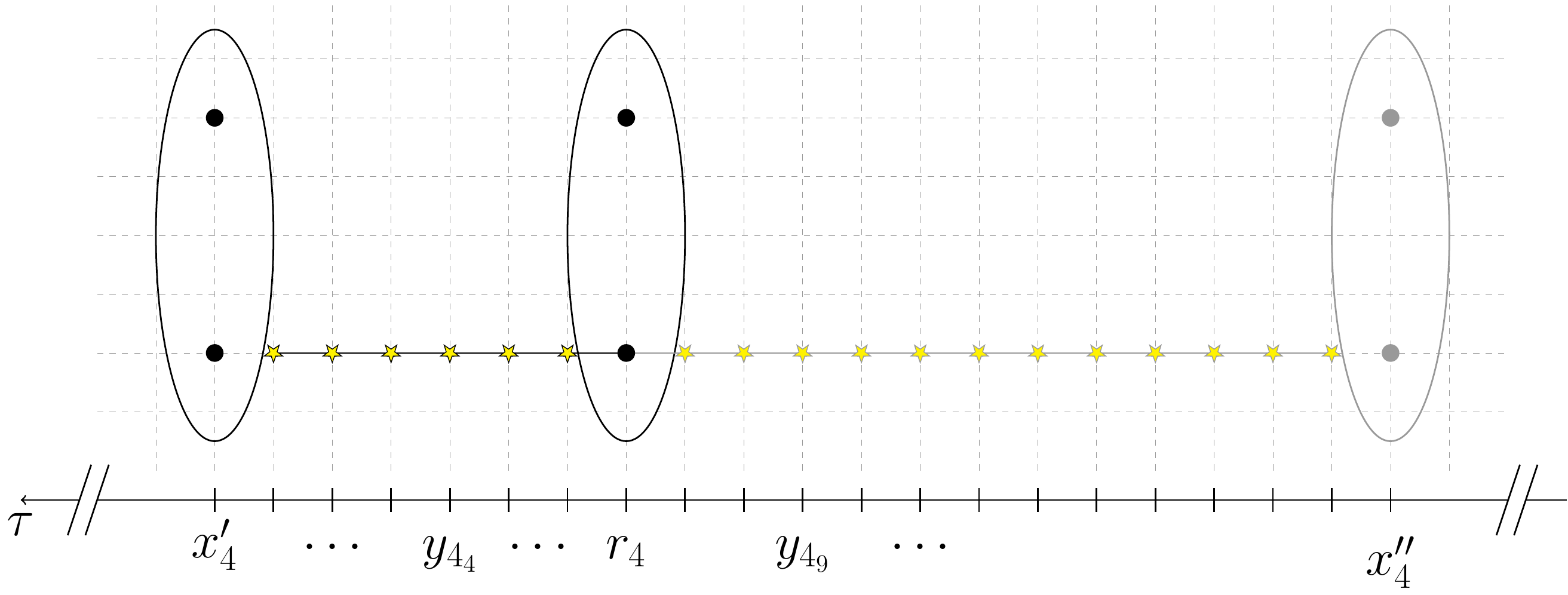}
  \caption{Parallelization of the insertion part of the three-point correlation function. Data from timeslices denoted with yellow stars is redistributed among all MPI ranks, 
such that all ranks have a similar workload.}
  \label{fig-2}
\end{figure}

The insertion part corresponds to the contraction of the stochastic propagator, i.e., the solution of the lattice Dirac equation sourced by random noise vectors, with the point-to-all propagator. This has to be repeated for each position $y_4$ of the insertion operator between the sinks~$x_4'$ and~$x_4$ and the source~$r_4$. Again, in order to keep a good workload balance we redistribute the data from the timeslices where the insertion is present, denoted with stars on figure \ref{fig-2}, among all MPI ranks in such a way that each rank has approximately the same number of insertion positions to work with, and we perform all the computations in parallel. Note that a separate Fourier transformation in Eq.~(\ref{insertion}) allows to select a desired momentum $\mathbf{q}$ flowing through the insertion.

\section{Performance}

We paid special attention to the data layout to enable the use of vectorization. The crutial point was to reorder the many loops
in the algorithm. We show this explicitly for pseudocode in listings \ref{alg-1} and \ref{alg-2}.
 
\begin{center}
\begin{minipage}[t]{5cm}
  \vspace{0pt}  
               \begin{algorithm}[H]
                    \For{{\color{blue}3 spin indices}}{
                    \For{{\color{green}color indices a,b}}{
                    \For{{\color{red}all sites in the local lattice}}{
                    \BlankLine
                    \BlankLine
                    }
                    }
                    }
                    \BlankLine
                    \BlankLine
                    \BlankLine
                    \BlankLine
                    \BlankLine
                    \BlankLine
                    \BlankLine
                    \caption{Reference implementation}
		    \label{alg-1}	
                \end{algorithm}
\end{minipage}%
\begin{minipage}[t]{5cm}
  \vspace{0pt}
                \begin{algorithm}[H]
                    \For{{\color{red}all sites in the local lattice}}{
                    \For{{\color{blue}1 spin index}}{
                    \For{{\color{green}color indices a,b}}{
                    \For{{\color{blue}2 spin indices SIMD vect.}}{
                    \BlankLine
                    \BlankLine
                    }
                    }
                    }
                    }
                    \BlankLine
                    \caption{LHA implementation}
	            \label{alg-2}
                \end{algorithm}

\end{minipage}
\end{center}

Benchmarks were performed on the QPACE3 supercomputer of the SFB/TR 55 at the Jülich Supercomputing Centre. The machine is based on Intel Xeon Phi (KNL) processors connected via Intel Omni-Path. We used the Intel compiler version 17.0.2.

The \texttt{LibHadronAnalysis} library \cite{lha:2017} was incorporated into the QDP++/Chroma software stack~\cite{Edwards:2004}, together with the multigrid solver \cite{richtmann:2016, wettig:2015, Frommer:2013, georg:2017, georg:2017qp3} optimized for the KNL architecture.

The computations were carried out on a single configuration of the CLS ensemble H101~\cite{Bruno:2015}. It is a $N_f=2+1$ ensemble with non-perturbatively $\mathcal{O}(a)$ improved Wilson fermions
and tree-level improved Symanzik gauge action and features open boundary conditions in time. The pion and kaon masses are about 420 MeV and this $32^3 \times 96$ lattice has a lattice spacing of about 0.086 fm.

Actual measurements were performed for the meson spectator part Eq.~(\ref{spectator}) and insertion part Eq.~(\ref{insertion}) using $r_4 = 30\,a$ and a source-sink separation of $r_4 - x'_4 = x''_4 - r_4 = 10 \, a$. Within the spectator part propagators are smeared at the source and the sink time-slice while in the insertion part only the source time-slice of the propagator is smeared. The number of stochastic indices is set to $N_i = 50$.

\begin{figure}[thb] % no figure before 1st section
  \centering
  \includegraphics[width=6cm,clip]{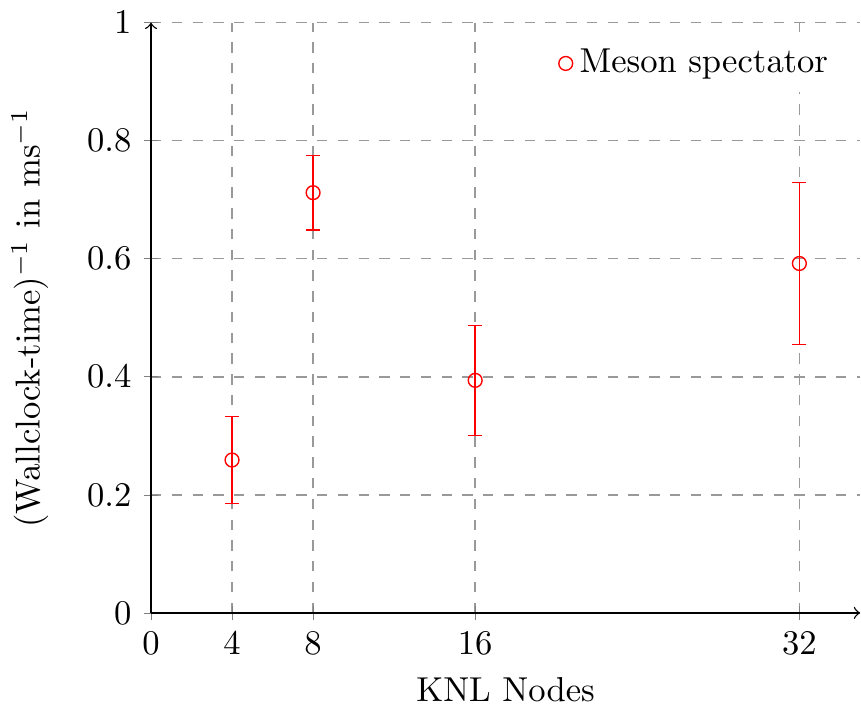}
  \includegraphics[width=6cm,clip]{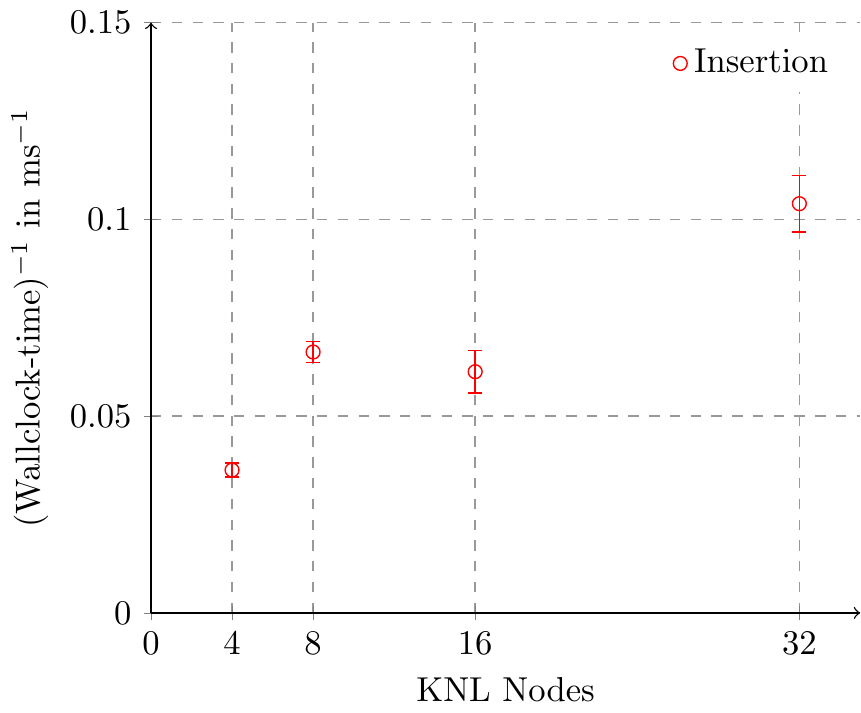}
  \caption{Strong scaling for the spectator (left) and the insertion (right) parts for the meson three-point correlation function. Measurements were done on a $32^3 \times 96$ lattice, with $\mathbf{k}^2 = \mathbf{k'}^2 = 0$ and 100 stochastic estimates for the stochastic propagators.}
  \label{fig-3}% Give a unique label
\end{figure}

Running on $4-32$ KNLs and distributing the $256$ hardware threads on each KNL to $8$ MPI tasks yields the strong scaling for the meson spectator and insertion part contraction as shown in figure~\ref{fig-3}. Note that the computation is done for $50$ independently seeded stochastic estimators in forward and backward direction, i.e., the mean values are averaged over $100$ computations each. The creation times for the noise and the solution vectors are not considered.

For the spectator/insertion a minimal computation time is achieved using $8/32$ KNLs with $8$ tasks per node. In this setup the Intel Omni-Path connection between the nodes becomes saturated, the computation is well parallelized and the overhead due to internal communication is not dominant. 

\begin{figure}[thb] % no figure before 1st section
  \centering
  \includegraphics[width=6cm,clip]{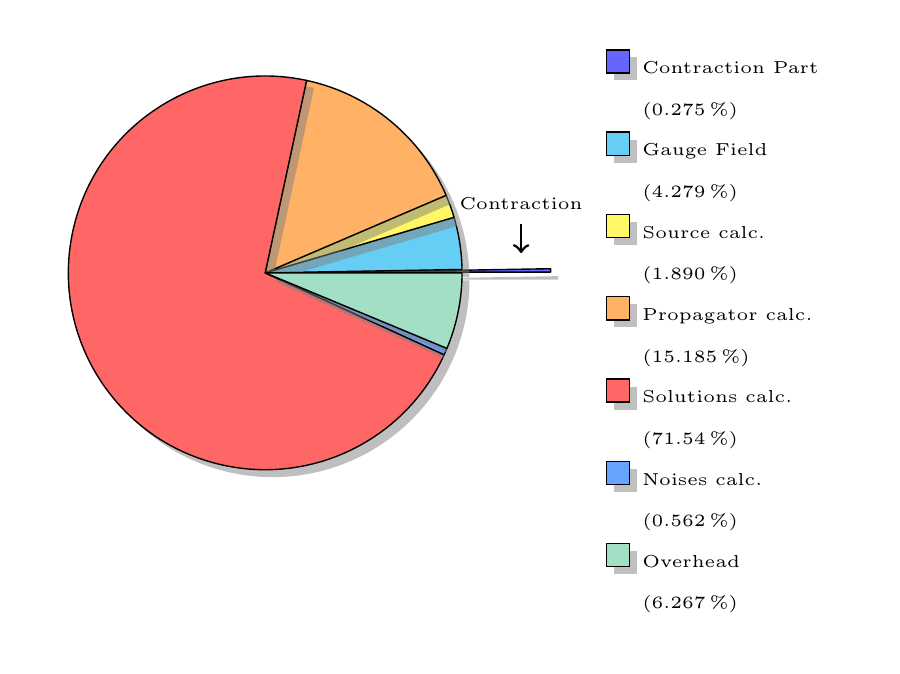}
  \includegraphics[width=6cm,clip]{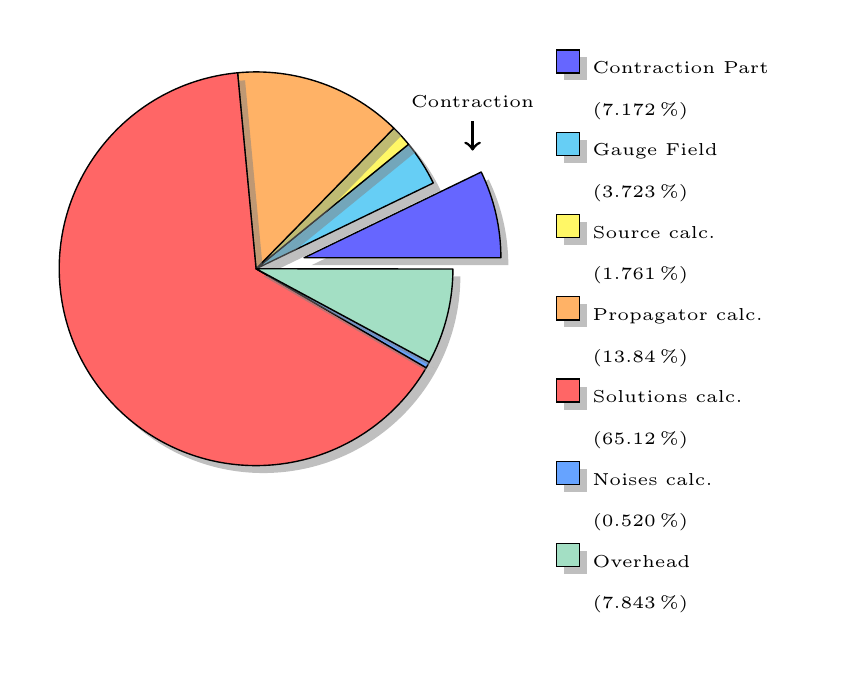}
  \caption{Contribution of the contraction time to the total time budget of the three-point correlation function estimation. On the left panel the case of total momentum 0 is shown, whereas on the right panel the spectator and insertion momentum square of $0, \dots, 8$ (i.e., $93 \cdot 93$ momentum combinations) is shown.}
  \label{fig-8}% Give a unique label
\end{figure}

In addition the wallclock-time for a particular measurement using two different ranges of momenta on one H101 configuration is evaluated -- again $8$ KNLs with $8$ tasks per node are used. The timings for $\mathbf{k'} = \mathbf{k}^2 = 0$ and for $\mathbf{k'}, \mathbf{k}^2 = 0, \dots, 8$ for spectator and insertion momenta are shown in figure \ref{fig-8}, where $p'_i = k'_i \, 2 \pi / L$, $q_i = k_i \, 2 \pi / L$ where the integers $k'_i$ and $k_i$ label the momentum components of $\mathbf{p}'$ and $\mathbf{q}$ within the Fourier transformation. Note that in these timings also a baryon measurement is included. In both cases the overall computation time is almost the same
\begin{itemize}
	\item[] \texttt{LibHadronAnalysis} wallclock-time for $\mathbf{k'}^2 = \mathbf{k}^2 = 0$: $\approx 530$s
	\item[] \texttt{LibHadronAnalysis} wallclock-time for $\mathbf{k'}^2 , \mathbf{k}^2 = 0,\dots, 8$ ($93 \cdot 93$ mom. combinations): $\approx 575$s.
\end{itemize}
Hence it is possible to produce data for a large number of final momenta without increasing the computation time significantly. In addition the data-layout presented in Eq.~(\ref{spectator}) and Eq.~(\ref{insertion}) provides analysis capabilities for various physical channels since the $\Gamma$-structures of the source and sink interpolators are not specified during the simulations.

The $\mathbf{k'}^2 , \mathbf{k}^2 = 0,\dots, 8$ measurement was also performed using the sequential source method \cite{martinelli:1989} which yields a run-time of $\approx 930$s. Compared to the above wallclock-time of $\approx 575$s this is a speed-up of $\approx 1.6$ where we have not taken into account that the stochastic code gives $16 \cdot 16$ $\Gamma$-combinations at source and sink for free. First tests have shown that the computation time of the analysis code needed to obtain final three-point function results is in the range of a few seconds and therefore is negligible. 

Collectively this means that one needs $\approx 82$ KNL core hours to perform the above measurement on a single configuration of the H101 assuming that 8 nodes with 64 cores per node are used. Altogether 2016 configurations are available to analyse the H101, i.e., at most $\approx 165000$ KNL core hours are needed to analyze the entire ensemble for every combination of source and sink meson and baryon interpolators on a single source time slice and for the given source sink separation $\Delta t = 10 \, a$. Due to the distribution shown in figure \ref{fig-8} the overall computation time should remain almost constant when increasing the number of source positions, always provided that the computation of the spectator part can be fully parallelized. 

\section{Conclusions}

The timings presented in the last section reveal that our implementation overtakes currently implemented methods at least by a factor of $1.5$ in terms of computation time for high momentum square for a given source-sink structure. However, exploiting the great flexibility of \texttt{LibHadronAnalysis} output data makes it feasible to reuse the data for many source-sink Dirac $\Gamma$-matrix combinations which results in an incredible economization. Furthermore it is sufficient to compute the insertion part only once and use it in both, baryon and meson, measurements. Since it is now possible to generate data containing an enormous amount of information it is also necessary to process the data further to finally get physical observables. The analysis software package is still in development and will be released as soon as possible.

\section*{Acknowledgements}
This work is funded by Deutsche Forschungsgemeinschaft (DFG) within the transregional collaborative
research centre 55 (SFB-TRR55). The Chroma software suite \cite{Edwards:2004} was used extensively
in this work along with the multigrid solver implementation of \cite{georg:2017qp3}. Computations
were performed on the SFB/TR55 QPACE supercomputers. 
%\clearpage
\bibliography{lattice2017}

%%%%%%%%%%%%%%%%%%%%%%%%%%%%%%%%%%%%%%%%%%%%%%%%%%%%%%%%%%%%%%%%%%%%%%%%%%%%%
\end{document}